\documentclass[preprintnumbers,amsmath,11pt,amssymb,floatfix,superscriptaddress,showpacs,nofootinbib]{revtex4}
\usepackage{graphicx}
\usepackage{epsfig}
\usepackage{bm}
\usepackage{amsfonts}

\input epsf

\begin{document}

\title{\Large Extended holographic Ricci dark energy in chameleon Brans-Dicke cosmology}

\author{Surajit Chattopadhyay}
\email{surajit_2008@yahoo.co.in, surajcha@iucaa.ernet.in}
\affiliation{ Pailan College of Management and Technology, Bengal
Pailan Park, Kolkata-700 104, India.}

\begin{abstract}
In the present work we have studied some features of the
generalized Brans-Dicke (BD) model in which the scalar field is
allowed to couple non-minimally with matter sector. Extended
holographic Ricci dark energy (EHRDE) has been considered in the
above framework of BD cosmology. Some restrictions have been
derived for the BD parameter $\omega$ and a stronger
matter-chameleon coupling has been observed with expansion of the
universe. In this framework, the equation of state parameter of
EHRDE has behaved like quintom. Also, we have reconstructed the
potential and coupling function for BD model for the EHRDE. It has
been observed that potential function is increasing as the
matter-chameleon coupling is getting stronger.
\end{abstract}


\maketitle

\section{Introduction}
Accumulating the observational data of Supernovae Type Ia (SN Ia)
by the year 1998, Riess et al. \cite{obs1} in the High-redshift
Supernova Search Team and Perlmutter et al. \cite{obs2} in the
Supernova Cosmology Project Team have independently reported that
the present universe is accelerating. The source for this
late-time acceleration was dubbed ``dark energy" (DE), which is
distinguished from ordinary matter species such as baryons and
radiation, in the sense that it has a negative pressure. Reviews
on DE are available in \cite{DE1,DE2,DE3,DE4,DE5,DE6,DE7}. The SN
Ia observations have shown that about $70\%$ of the present energy
of the universe consist of DE \cite{DE5}. The simplest candidate
for DE is the so-called cosmological constant $\Lambda$, whose
energy density remains constant \cite{cosmoconst}. If the origin
of DE is not the cosmological constant, then one may seek for some
alternative models for DE to explain the cosmic acceleration
today. There are two approaches to construct models of DE other
than cosmological constant \cite{DE5}: (i) To modify the right
hand side of the Einstein equations by considering specific forms
of energy-momentum tensor $T_{\mu\nu}$ with negative pressure;
(ii) To modify the left hand side of the Einstein equations. The
representative models that belong to the first category are the
so-called quintessence \cite{quintessence}, k-essence
\cite{kessence} and perfect fluid models \cite{perfectfluid}. The
representative models that belong to the second class are
``modified gravity"  \cite{mod1,mod2,mod3,mod4} models that
include $f(R)$ gravity \cite{fr}, scalar-tensor theories
\cite{scalartensor}, and braneworld models \cite{brane}. An
extensive review on modified gravity is presented in \cite{mod5}.
Among all scalar-tensor theories of gravity, the simplest one is
the so-called \emph{Brans-Dicke theory} (BD) introduced by Brans
and Dicke \cite{BD} that modifies general relativity in accordance
with Mach's principle. Cosmological models of classical BD theory
were first studied in \cite{BD1,BD2}. In the subsequent decades,
various aspects of BD cosmology have been widely investigated
\cite{BD3,BD4,BD5,BD6,BD7,BD8,BD9,BD10}.

The present paper is aimed at working in chameleon Brans-Dicke
cosmology. Thus, we discuss a bit of chameleon cosmology in this
place. Khoury and Weltman \cite{ch1,ch2} employed
self-interactions of the scalar field to avoid the bounds on such
a field and dubbed such scalars to be ``chameleon fields" due to
the way in which the field's mass depends on the density of matter
in the local environment. In regions of high density, the
chameleon ``blends" with its environment and becomes essentially
invisible to searches for violation of equivalence principle and
the fifth force \cite{ch1}. Ref. \cite{ch3} have shown that
chameleon scalar field can provide explicit realizations of a
quintessence model where the quintessence scalar field couples
directly to baryons and dark matter with gravitational strength.
Ref. \cite{ch4} have illustrated that interacting chameleon field
plays an important role in late time universe acceleration and
phantom crossing. Ref. \cite{ch5} proposed an interacting
holographic dark energy model in chameleon-tachyon cosmology by
interaction between the components of the dark sectors. Based on
two independent functions of the scalar field, ref. \cite{ch6}
constructed an exact solution describing the evolution of the type
Bang-to-Rip with the phantom divide line crossing in the chameleon
cosmology.

Now we come to the chameleon Brans-Dicke cosmology. In the models,
where a non-minimal coupling between the scalar field and matter
system is considered by introducing an arbitrary functions of the
scalar field, the scalar field is regarded as as a chameleon field
\cite{BD10}. Ref. \cite{chbd1} were the first to introduce this
kind of chameleon-matter coupling in the BD model to achieve
accelerated expansion of the universe. In a recent paper, ref.
\cite{BD10} considered a generalized BD model with power-law form
of the scale factor and the coupling functions as the inputs and
allowed non-minimal coupling with the matter sector to show that
accelerated expansion of the universe can be realized for a
contrained range of exponents of the potential function. Treating
the BD scalar field as a chameleon scalar field and taking a
non-minimal coupling of the scalar field with matter, ref.
\cite{setare1} studied cosmological implication of holographic
dark energy in the BD gravity. Ref. \cite{jamil1} investigated the
cosmological applications of interacting holographic dark energy
in BD theory with chameleon scalar field which is non-minimally
coupled to the matter field and observed that phantom crossing can
be constructed if the model parameters are chosen suitably. For
different epochs of the cosmic evolution, ref. \cite{jamil2}
investigated the BD chameleon theory of gravity and obtained exact
solutions of the scale factor, scalar field and potential. Ref.
\cite{nabushi} discussed late-time dynamics of a chameleonic
generalized BD cosmology with the power law chameleonic function
$f(\phi)\propto \phi^n$, where $n$ is a real parameter motivated
from string theories.

Making an approach to dark energy in the frame of quantum gravity
is the well-known holographic dark energy (HDE) inspired by the
holographic principle \cite{HDE1}. The HDE, whose density is
$\rho_{\Lambda}=3c^2M_p^2L^{-2}$ was proposed by Li \cite{HDE2}.
Subsequent studies on HDE from various points of view include
\cite{HDE3,HDE4,HDE5,HDE6}. In the present work we are considering
a special form of HDE \cite{EHRDE1} dubbed as ``extended
holographic Ricci dark energy" (EHRDE) \cite{EHRDE2}, whose
density has the form
\begin{equation}\label{EHRDE}
\rho_{\Lambda}=3M_p^2(\alpha H^2+\beta \dot{H})
\end{equation}
where, $M_p^2$ is the reduced Planck mass, $\alpha$ and $\beta$
are constants to be determined. Like ref. \cite{setare1}, we shall
assume that $\rho_{\Lambda}$ and pressureless dark matter are
conserved separately and we shall assume a non-minimal coupling
between the scalar field and the matter field.
\\
\section{EHRDE in chameleon BD cosmology}
We begin with the chameleon BD theory in which the scalar field is
coupled non-minimally to the matter field via the action
\cite{BD10}
\begin{equation}\label{action}
S=\frac{1}{2}\int d^4x\sqrt{-g}\left(\phi
R-\frac{\omega}{\phi}g^{\mu\nu}\nabla_{\mu}\phi\nabla_{\nu}\phi-2V+2f(\phi)L_m\right)
\end{equation}
In equation (\ref{action}), $R$ and $\phi$ denote the Ricci scalar
and BD scalar field respectively. The $f(\phi)$ and $V(\phi)$ are
analytic functions of the scalar field. The matter Lagrangian
density, denoted by $L_m$, is coupled with $\phi$ via the function
$f(\phi)$. This function allows a non-minimal coupling between the
matter system and the scalar field. If $f(\phi)=1$, we get back
the BD action with potential function $\phi$ \cite{BD10}. Varying
action with respect to the metric $g_{\mu\nu}$ and $\phi$ one gets
the field equations \cite{BD10}
\begin{equation}\label{field1}
\phi G_{\mu\nu}=T_{\mu\nu}^\phi+f(\phi)T_{\mu\nu}^m
\end{equation}
\begin{equation}\label{field2}
(2\omega+3)\Box \phi+2(2V-V'\phi)=T^m f-2f'\phi_m
\end{equation}
where $\Box=\nabla^{\mu}\nabla_\mu$~, where $\nabla_\mu$
represents the covariant derivative, $T^m=g^{\mu\nu}T_{\mu\nu}^m$
and the prime denotes differentiation with respect to $\phi$. In
equation (\ref{field1})
\begin{equation}\label{field11}
T_{\mu\nu}^\phi=\frac{\omega}{\phi}\left(\nabla_{\mu}\phi
\nabla_{\nu}\phi-\frac{1}{2}g_{\mu\nu}\nabla_{\alpha}\phi\nabla^{\alpha}\phi\right)+\left(\nabla_{\mu}\nabla_{\nu}\phi-g_{\mu\nu}\Box
 \phi \right)-V(\phi)g_{\mu\nu}
\end{equation}
and
\begin{equation}\label{field12}
T_{\mu\nu}^m=\frac{-2}{\sqrt{-g}}\frac{\delta(\sqrt{-g}L_m)}{\delta
g^{\mu\nu}}
\end{equation}
Because of the explicit coupling between matter system and $\phi$
the stress tensor $T_{\mu\nu}^m$ is not divergence free.

Now we shall apply the above framework to a homogeneous and
isotropic universe described by the Friedman-Robertson-Walker
metric
\begin{equation}\label{FRW}
ds^2=-dt^2+a^2(t)\left(\frac{dr^2}{1-kr^2}+r^2 d\Omega^2\right)
\end{equation}
The universe is open, closed or flat according as
$k=-1,~+1~\textrm{or}~0 $. In a spatially flat universe, equations
(\ref{field1}) and (\ref{field12}) yield
\begin{equation}\label{fried1}
3H^2=\frac{f}{\phi}\rho+\frac{\omega}{2}\frac{\dot{\phi}^2}{\phi^2}-3H
\frac{\dot{\phi}}{\phi}+\frac{V}{\phi}
\end{equation}
\begin{equation}\label{fried2}
3(\dot{H}+H^2)=-\frac{3\rho}{\phi(2\omega+3)}\left[\gamma \phi
f'+\left(\omega\left(\gamma+\frac{1}{3}\right)+1\right)f\right]-\omega\frac{\dot{\phi}^2}{\phi^2}+3H
\frac{\dot{\phi}}{\phi}+\frac{1}{2\omega+3}\left[3V'+(2\omega-3)\frac{V}{\phi}\right]
\end{equation}
\begin{equation}\label{fried3}
(2\omega+3)(\ddot{\phi}+3H\dot{\phi})-2(2V-\phi
V')=\rho\left((1-3\gamma)f+2\gamma\phi f'\right)
\end{equation}
In equations (\ref{fried1}) to (\ref{fried3}),
$\rho=\rho_m+\rho_\Lambda$ and
$\gamma=\frac{p_{\Lambda}}{\rho_m+\rho_\Lambda}$ (since dark
matter is pressureless, $p_m=0$). This approach is similar to
refs. \cite{setare1,jamil1}.

In the present work, we shall investigate two cases. In one case
we shall assume special forms for $V$, $f$, $a$ and $\phi$ and
derive conditions that strengthen matter-chameleon coupling with
expansion of the universe. In another case we shall not assume any
form for $V$ and $f$. Rather we shall reconstruct them for the
EHRDE.

\section{Discussion}
\subsection{Case I}
In this section we shall consider a set of ansatz for the
potential function, analytic function, scalar field and scale
factor. Based on them, we shall determine constraint on the BD
parameter $\omega$. Subsequently, based on the constraints on
$\omega$ and other parameters we shall investigate the behavior of
the EHRDE in BD chameleon cosmology. Following \cite{BD10} we
choose
\begin{equation}\label{assumption}
V(\phi)=V_0
\phi^{l_1}~;f(\phi)=f_0\phi^{l_2};~a(t)=a_0t^n;~\textrm{and}~\phi(t)=\phi_0t^m
\end{equation}

Since $M_p^2=1/8\pi G$ and in BD theory $\phi\propto G^{-1}$
\cite{jamil1} we can take EHRDE in the chameleon BD as
\begin{equation}\label{densityBD}
\rho_{\Lambda}=3\phi(\alpha H^2+\beta \dot{H})
\end{equation}
and using conservation equation
$\dot{\rho}_{\Lambda}+3H(\rho_{\Lambda}+p_{\Lambda})=0$ we get
\begin{equation}\label{pressureBD}
p_{\Lambda}=\phi\left((1-3
H^2-2\dot{H})\alpha+\frac{\beta}{H^2}\left(\dot{H}-H(\ddot{H}+3H\dot{H})\right)\right)
\end{equation}
Using equations (\ref{assumption}) - (\ref{pressureBD}) in
(\ref{fried1}) we have
\begin{equation}\label{condition}
\varrho_1 t^{ml_2-m+2-3n}+\varrho_2t^{2-m+ml_1}+\varrho_3
t^{ml_2}=\varrho_4
\end{equation}
where, $\varrho_1,~\varrho_2,~\varrho_3~\textrm{and}~\varrho_4$
are terms involving constant terms of ansatz (\ref{assumption}).
In equation (\ref{condition}) we observe that the right hand side
is constant with $t$ involved in the left hand side. This is
possible if the powers of $t$ are $0$. In (\ref{condition})
$ml_2\neq 0$. However, its coefficient
$3nf_0\phi_0^{l_2}\left(n\alpha-\beta\right)(3+\omega+3\gamma(l_2+\omega))$
can be set equal to $0$. Since $n\alpha-\beta=0$ will lead to
$p_{\Lambda}=0$, we can set $3+\omega+3\gamma(l_2+\omega)=0$.
Finally, we get the restrictions
\begin{equation}\label{restriction1}
\omega=-\frac{3(1+\gamma l_2)}{1+3 \gamma}
\end{equation}
\begin{equation}\label{restriction2}
-2=m(l_2-1)-3n
\end{equation}
\begin{equation}\label{restriction3}
-2=m(l_1-1)
\end{equation}
The above restrictions are also found valid if we use the ansatz
(\ref{assumption}) and (\ref{densityBD}) - (\ref{pressureBD}) in
(\ref{fried2}). In (\ref{restriction1}), we have for accelerated
expansion of the universe $3\gamma+1>0$. Hence,
(\ref{restriction1}) finally leads to a constraint on the BD
parameter as
\begin{equation}\label{restriction11}
\omega>\frac{l_2-3}{1+3\gamma}
\end{equation}
 In 1973 $\omega > 5$ was consistent with known data. By 1981 $\omega > 30$ was consistent with known data.
 In 2003, evidence derived from the Cassini-Huygens experiment shows that the value of $\omega$ must exceed $40,000$
 \cite{Wicke}.Hence, we can rewrite
\begin{equation}\label{restriction111}
\omega>\frac{l_2-3}{1+3\gamma}\gg 1
\end{equation}
Hence, $l_2> 4$. From (\ref{restriction2}) and
(\ref{restriction3}) we have
\begin{equation}\label{n}
n=\frac{2}{3}\left(\frac{l_1-l_2}{l_1-1}\right);~~\textrm{and}~~m=\frac{2}{1-l_1}
\end{equation}
If $l_1>1$ then $m<0$, which implies that the scalar field is a
decreasing function of $t$. Since $n$ has to be positive,
$l_1>l_2$ from (\ref{n}). Again, if $l_1<1$, then $m>0$. Since
$l_2>0$, $ml_2>0$. Thus, for $l_1<1$, coupling function $f\propto
t^{ml_2}$ increases with time i.e. matter-chameleon coupling gets
stronger as the universe expands.

Choosing appropriate parameters (keeping the restrictions in mind)
we have plotted the EoS parameter $\gamma=\frac{p}{\rho}$ for
different combinations of parameters in Fig. 1. The EoS parameter
clearly shows a transition from $\gamma>-1$ (quintessence) to
$\gamma<-1$ (phantom). Hence, the EoS parameter is found to
exhibit ``quintom"-like behavior. Next we plot in Fig. 2 the
deceleration parameter $q$ \cite{EHRDE2}
\begin{equation}\label{q}
q=\frac{1}{2}+\frac{3p}{2\rho}
\end{equation}
We observe that the deceleration parameter has a clear transition
from positive to negative side. This indicates the transition from
decelerated to accelerated phase of the universe.
\begin{figure}[h]
\includegraphics[width=16pc]{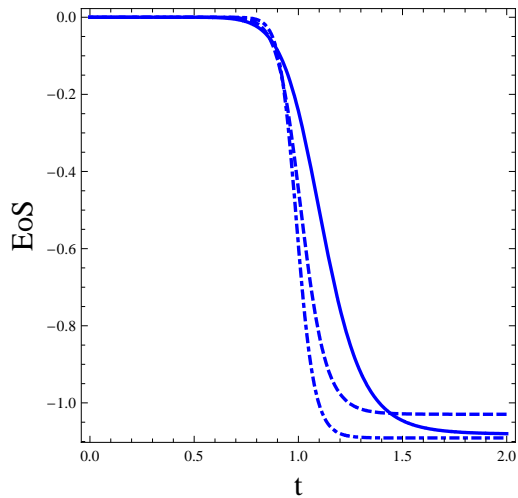}
\caption{\label{label}Evolution of EoS parameter $\gamma$ against
$t$. The solid, dot-dashed and dashed lines have the parameter
combinations $\alpha=2.3,1.3,1.5$,~
$\beta=4.5,2.5,1.5$,~$l_1=0.3,0.5,0.2$ and $l_2=4.5,6,7$.}
\end{figure}
\begin{figure}
\includegraphics[width=16pc]{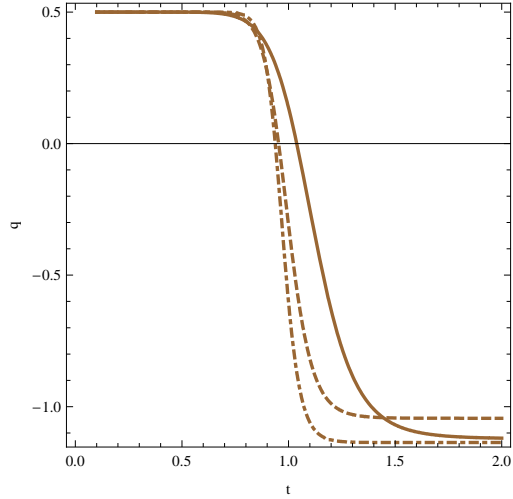}
\caption{\label{label}Evolution of deceleration parameter $q$
(equation (\ref{q})) against $t$. The solid, dot-dashed and dashed
lines have the parameter combinations $\alpha=2.3,1.3,1.5$,~
$\beta=4.5,2.5,1.5$,~$l_1=0.3,0.5,0.2$ and $l_2=4.5,6,7$.}
\end{figure}

\subsection{Case II}
In this section we shall only assume
$a(t)=a_0t^n;~\textrm{and}~\phi(t)=\phi_0t^m$. We shall use these
ansatz to reconstruct the potential $V$ and the analytic function
$f$ of the chameleon BD cosmology. On putting these ansatz in
equation (\ref{fried1}) we get an equations that involves $V$, $f$
and remaining terms as functions of $t$
\begin{equation}\label{V}
V=-\frac{1}{a_{0}^{3}}\rho_{m0}t^{-3n}f+\frac{\phi_{0}t^{m-2}}{2}\left[-m^2\omega+6n(m+n-(n-\beta)f)\right]
\end{equation}
We again differentiate it with respect to $t$ and get a new
equation involving $\dot{V}$ and $\dot{f}$. Using this in
(\ref{fried2}) along with the said ansatz we get the following
differential equation on $f$ with $t$ as the independent variable.
\begin{equation}\label{DE}
\frac{df(t)}{dt}-\frac{3(m+2n)}{2t}f(t)=\frac{3a_0^3\phi_0t^{m+3n-1}\left(mn(t(3+2\omega)-4-2n)+m^3(1+\omega)-m^2(1+2n)-4n^2\right)}{2\left(\rho_{m0}t^2-a_0^3(m-2)\phi_0(n\alpha-\beta)\right)}
\end{equation}
Solving (\ref{DE}) we get reconstructed analytic function
$f(\phi)$ as
\begin{equation}
\begin{array}{c}\label{f}
f(\phi)=f(\phi_0 t^m)=t^{m+3n}\left[C_{1}t^{\frac{m}{2}}-\frac{2}{\rho_{m0}(8+6m+m^2)t^2}\right.\\
\left.\left\{mn(4+m)(3+2\omega)t~\textrm{Hypergeometric}2F1\left[\frac{2+m}{4-2m-6n},1,\frac{6(n-1)+m}{2(3n+m-2)},\frac{a_0^3(m-2)\phi_0(n\alpha-\beta)}{\rho_{m0}}\right]\right.\right.\\
\left.\left.+(2+m)(-4n^2-2mn(2+n)-m^2(1+2n)+m^3(1+\omega))\right.\right.\\
\left.\left.\times \textrm{Hypergeometric}2F1\left[\frac{4+m}{4-2m-6n},1,\frac{6n-8+m}{2(3n+m-2)},\frac{a_0^3(m-2)\phi_0(n\alpha-\beta)}{\rho_{m0}}\right]\right\}\right]\\
\end{array}
\end{equation}
Using (\ref{f}) in (\ref{V}) we get the reconstructed potential
function as
\begin{equation}\label{Vsoln}
\begin{array}{c}
V(\phi)=V(\phi_0t^m)=\frac{t^{-4+m}}{2a_0^3(2+m)(4+m)\rho_{m0}}\times~~~~~~~~~~~~~~~~~~~~~~~~~~~~~~~~~~~~~~~~~~~~~~~~~~~~\\
\left(-(2+m)(4+m)\rho_{m0}t^2(2C_1\rho_{m0}t^{\frac{4+m}{2}}+a_0^3\phi_0(-6mn+6n(-n+C1t^{\frac{3}{2}(m+2n)}(n\alpha-\beta))+m^2\omega))\right)~~~~~~~~~~~~~~~~~~~~~~~~~~~~~~~~~~~~~~~~~~~~~~~~~~~~~~~~~~~~~~~~~~\\
+4(\rho_{m0}t^2+3a_0^3n\phi_0t^{m+3n}(n\alpha-\beta))\times~~~~~~~~~~~~~~~~~~~~~~~~~~~~~~~~~~~~~~~~~~~~~~~~~~~~~~~~~~~~~~~~~~~\\
\left[mn(4+m)(3+2\omega)t~\textrm{Hypergeometric}2F1\left[1,\frac{2+m}{4-2m-6n},1+\frac{2+m}{4-2m-6n},\frac{a_0^3(m-2)\phi_0(n\alpha-\beta)t^{-2+m+3n}}{\rho_{m0}}\right]+\right.~~~~~~~~~~~~~~~~~~~~~~~~~~~~~~~~~~~~~~~~~~~~~~~~~~~~~~~~~~~~~\\
\left.(2+m)(-4n^2-2mn(2+n)-m^2(2n+1)+m^2(1+\omega))~~~~~~~~~~~~~~~~~~~~~~~~~~~~~~~~~~~~~~~~~~~~~~~~~~~~~~~~~~~~~~~~~\right.\\
\left.\times\textrm{Hypergeometric}2F1\left[1,\frac{4+m}{4-2m-6n},1+\frac{4+m}{4-2m-6n},\frac{a_0^3(m-2)\phi_0(n\alpha-\beta)t^{-2+m+3n}}{\rho_{m0}}\right]\right]~~~~~~~~~~~~~~~~~~~~~~~~~~~~~~~~~~~~~~~~~~~~~~~~~~~\\
\end{array}
\end{equation}

\begin{figure}[h]
\includegraphics[width=16pc]{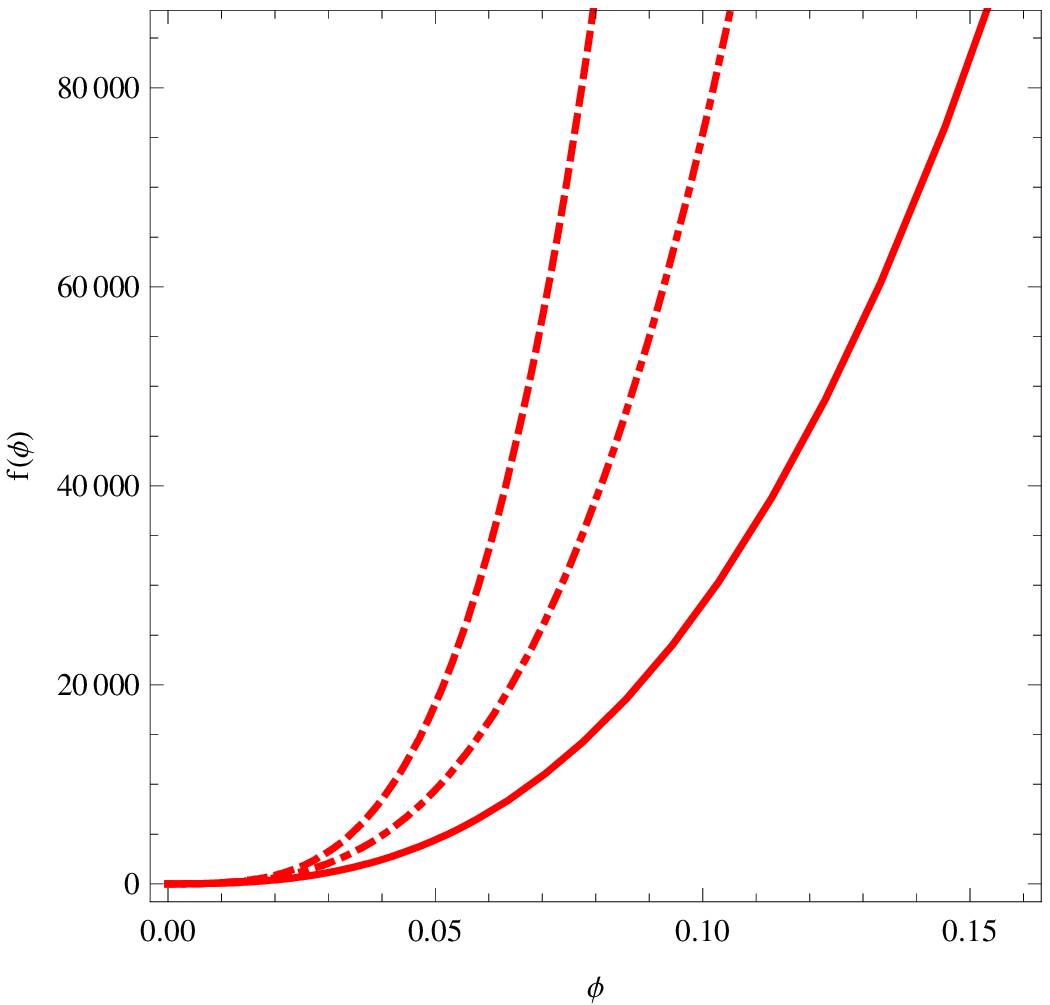}
\caption{\label{label}Reconstructed $f(\phi)$ against $\phi$ for
various combination of parameters.}
\end{figure}

\begin{figure}[h]
\includegraphics[width=16pc]{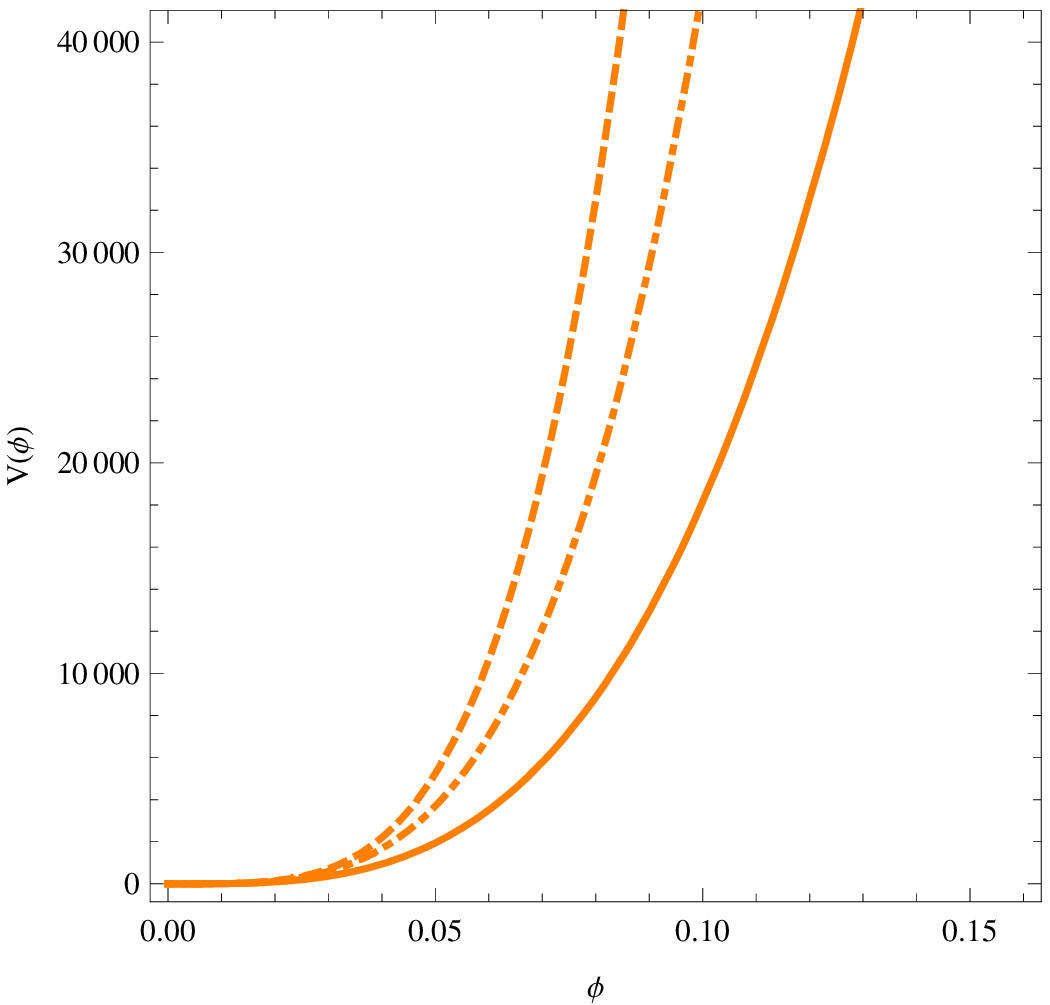}
\caption{\label{label}Reconstructed $V(\phi)$ against $\phi$ for
various combination of parameters.}
\end{figure}

\begin{figure}[h]
\includegraphics[width=16pc]{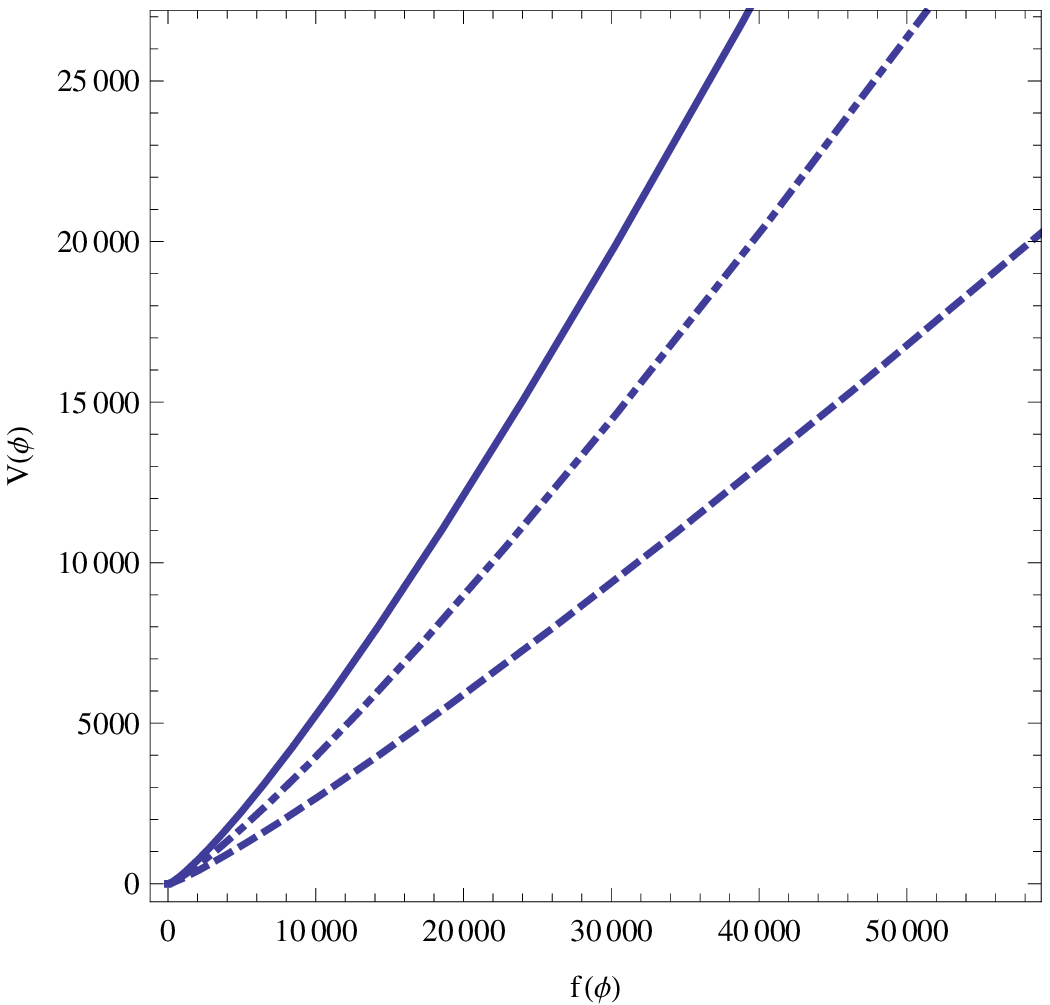}
\caption{\label{label}Reconstructed $V(\phi)$ against $f(\phi)$
for various combination of parameters.}
\end{figure}

\begin{figure}[h]
\includegraphics[width=16pc]{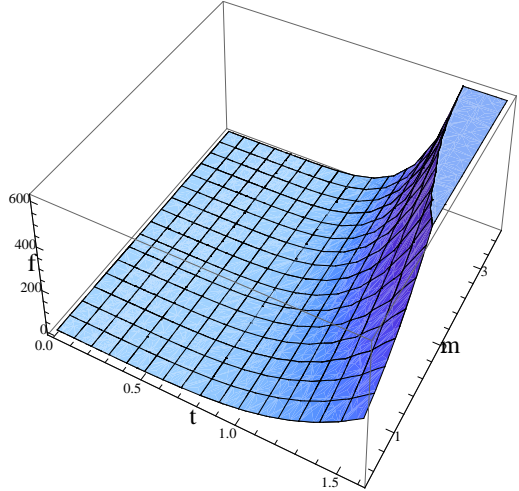}
\caption{\label{label}Reconstructed $f(\phi)$ against $t$ for a
range of values of $m$.}
\end{figure}

In Figs. 3, 4 and 5 we have plotted the reconstructed $f(\phi)$
against $\phi$, reconstructed $V(\phi)$ against $\phi$ and
reconstructed $V(\phi)$ against $f(\phi)$ respectively. We observe
that $f(\phi)\rightarrow 0$ as $\phi\rightarrow 0$ and
$V(\phi)\rightarrow 0$ as $\phi\rightarrow 0$. These indicate
satisfaction of one of the sufficient conditions for realistic
models. In Fig. 5 we observe that the potential function $V(\phi)$
is increasing with the coupling function $f(\phi)$. Thus, we infer
that as the matter-chameleon coupling gets stronger, the potential
function also increases. In Fig. 6 we have plotted $f(\phi)$
against $t$ for a range of positive values of $m$. This figure
shows that $f(\phi)$ in an increasing function of the cosmic time
$t$. This indicates that with the expansion of the universe, the
matter-chameleon coupling gets stronger, where the coupling
function has been reconstructed for EHRDE.
\\
\section{Conclusion}
In the present work we have studied some features of the
generalized BD model in which the scalar field is allowed to
couple non-minimally with matter sector. Extended holographic
Ricci dark energy has been considered in the above framework of BD
cosmology. The energy density has been considered in the form
$\rho_{\Lambda}=3\phi(\alpha H^2+\beta \dot{H})$ to accommodate
the BD scalar field in the dark energy density. The work has been
carried out in two different perspectives. Firstly, we have
assumed the ansatz $V(\phi)=V_0
\phi^{l_1}~;f(\phi)=f_0\phi^{l_2};~a(t)=a_0t^n;~\textrm{and}~\phi(t)=\phi_0t^m
$ and derived some restrictions on the BD parameter $\omega$ Under
this restriction we have investigated the equation of state
parameter for the dark energy and a quintom-like behavior has been
observed. Furthermore, the deceleration parameter has shown a
transition from decelerated to accelerated phase of the universe.
In this scenario, it has been observed that the restrictions are
leading to a stronger matter-chameleon coupling with the expansion
of the universe. Secondly, we have considered only
$a(t)=a_0t^n;~\textrm{and}~\phi(t)=\phi_0t^m$ and reconstructed
the coupling function and potential for the extended holographic
Ricci dark energy in BD model. We have observed that the potential
is increasing with increase in matter-chameleon coupling.
Furthermore, we have seen that for any positive value of $m$,
$f(\phi)$ is an increasing function of $t$. This means that for
this reconstruction, the coupling function is increasing i.e. the
matter-chameleon coupling is getting stronger with passage of
cosmic time.
\\
\textbf{Acknowledgement}\\
 The author wishes to acknowledge the
financial support from the Department of Science and Technology,
Govt. of India under the Fast Track Scheme for Young Scientists.
The Grant No. is SR/FTP/PS-167/2011. Also, Visiting Associateship
provided by the Inter-University Centre for Astronomy and Astrophysics (IUCAA), Pune, India is duly acknowledged.
\\


\begin{thebibliography}{99}

\bibitem{obs1} A. G. Riess \emph{et al}., \emph{Astron. J.} \textbf{116}
1009 (1998).
\bibitem{obs2} S. Perlmutter \emph{et al}., \emph{Astrophys. J.} \textbf{517} 565 (1999)
\bibitem{DE1} P. J. E. Peebles, B. Ratra \emph{Rev. Mod. Phys.} \textbf{75} 559
(2003).
\bibitem{DE2} E. J. Copeland, M. Sami and S. Tsujikawa \emph{Int. J. Mod. Phys. D} \textbf{15} 1753 (2006).
\bibitem{DE3} T. Padmanabhan \emph{Curr. Sci.} \textbf{88} 1057
(2005).
\bibitem{DE4} T. Padmanabhan \emph{Gen. Rel. Grav.} \textbf{40} 529
(2008).
\bibitem{DE5} L. Amendola, S. Tsujikawa \emph{Dark energy: theory and observations}. Cambridge University Press (2010).
\bibitem{DE6} K. Bamba, S. Capozziello, S. I. Nojiri, S. D. Odintsov \emph{Astrophys. Space Sci.} \textbf{342} 155 (2012).
\bibitem{DE7} J. Yoo, Y. Watanabe  \emph{Int. J. Mod. Phys. D} \textbf{21} 1230002
(2012).
\bibitem{cosmoconst} S. Weinberg \emph{Rev. Mod. Phys. A}
\textbf{23} 4273 (2008).
\bibitem{quintessence} S. M. Carroll \emph{Phys. Rev. Lett.}
\textbf{81} 3067 (1998).
\bibitem{kessence} T. Chiba, T. Okabe and M. Yamaguchi \emph{Phys. Rev.
D} \textbf{62} 023511 (2000).
\bibitem{perfectfluid} M. C. Bento, O. Bertolami and A. A. Sen
\emph{Phys. Rev. D} \textbf{66} 043507 (2002).
\bibitem{mod0} S. Nojiri and S. D. Odintsov \emph{Int. J. Geom. Meth. Mod. Phys.} \textbf{4} 115
(2007).
\bibitem{mod1} R. Myrzakulov \emph{Eur. Phys. J. C} \textbf{71}
1752 (2011).
\bibitem{mod2} E. Elizalde, R. Myrzakulov, V. V. Obukhov and D.
Sáez-Gómez \emph{Class. Quantum Grav.} \textbf{27} 095007 (2010).
\bibitem{mod3} K. Bamba, R. Myrzakulov, S. Nojiri and S. D.
Odintsov \emph{Phys. Rev. D} \textbf{85} 104036 (2012).
\bibitem{mod4} R. Myrzakulov \emph{Eur. Phys. J. C} \textbf{72} 2203 (2012).
\bibitem{fr} S. Capozzielo \emph{Int. J. Mod. Phys. D} \textbf{11}
483 (2002).
\bibitem{mod5} S. Nojiri and S. D. Odintsov \emph{Phys. Rep.}
\textbf{505} 59 (2011).
\bibitem{scalartensor} L. Amendola \emph{Phys. Rev. D} \textbf{60}
043501 (1999).
\bibitem{brane} V. Sahni and Y. Shtanov \emph{JCAP} \textbf{0311}
014 (2003).
\bibitem{BD} C. Brans and R. H. Dicke \emph{Phys. Rev.}
\textbf{124} 925 (1961).
\bibitem{BD1} G. S. Greenstein \emph{Astrophys. Lett.} \textbf{1} \textbf{139} (1968).
\bibitem{BD2} G. S. Greenstein \emph{Astrophys. Space Sci.} \textbf{2} 155 (1968).
\bibitem{BD3} M. K. Mak and T. Harko \emph{Europhys. Lett.} \textbf{60} 155
(2002).
\bibitem{BD4} A. E. Montenegro Jr and S. Carneiro \emph{Class. Quantum Grav.} \textbf{24} 313
(2007).
\bibitem{BD5} M. Arik and M. C. Calik \emph{JCAP} \textbf{01} 013
(2005).
\bibitem{BD6} S. Sen and A. A. Sen \emph{Phys. Rev. D} \textbf{63}
124006 (2001).
\bibitem{BD7} T. R. Seshadri \emph{J. Astrophys. Astr.} \textbf{18} 339
(1997).
\bibitem{BD8} S. Sen and T. R. Seshadri \emph{Int. J. Mod. Phys.
D} \textbf{12} 445 (2003).
\bibitem{BD9} Y. Bisabr \emph{Gen. Relativ. Gravit.} \textbf{44} 427
(2012).
\bibitem{BD10} Y. Bisbar \emph{Phys. Rev. D} \textbf{86} 127503
(2012).
\bibitem{ch1} J. Khoury and A. Weltman \emph{Phys. Rev. D} \textbf{69} 044026
(2004).
\bibitem{ch2} J. Khoury and A. Weltman \emph{Phys. Rev. Lett.} \textbf{93} 171104
(2004).
\bibitem{ch3} P. Brax, C. van de Bruck, A-C. Davis, J. Khoury and A.
Weltman \emph{Phys. Rev. D} \textbf{70} 123518 (2004).
\bibitem{ch4} H. Farajollahi and A. Salehi \emph{Astrophys. Space Sci.} \textbf{338} 375
(2012).
\bibitem{ch5} H. Farajollahi, A. Ravanpak  and G. F. Fadakar \emph{Astrophys. Space Sci.} \textbf{336}
461 (2011).
\bibitem{ch6} F. Cannata and A. Y. Kamenshchikannata \emph{Int. J. Mod. Phys. D} \textbf{20} 121
(2011).
\bibitem{chbd1} S. Das and N. Banerjee \emph{Phys. Rev. D} \textbf{78} 043512
(2008).
\bibitem{setare1} M. R. Setare and M. Jamil \emph{Phys. Lett. B}
\textbf{690} 1 (2010).
\bibitem{jamil1} A. Sheykhi and M. Jamil \emph{Phys. Lett. B}
\textbf{694} 248 (2011).
\bibitem{jamil2} M. Jamil, I. Hussain and D. Momeni \emph{Eur. Phys. J. Plus} \textbf{126} 80
(2011).
\bibitem{nabushi} R. A. El-nabulsi, \emph{Eur. Phys. J. Plus} \textbf{127} 23
(2012).
\bibitem{HDE1} R. Bousso \emph{Rev. Mod. Phys.} \textbf{74} 825
(2002).
\bibitem{HDE2} M. Li \emph{Phys. Lett. B} \textbf{603} 1 (2004).
\bibitem{HDE3} M. R. Setare and M. Jamil \emph{JCAP} \textbf{02} 010
(2010).
\bibitem{HDE4}  M. Jamil, M. U. Farooq and M. A. Rashid \emph{Eur. Phys. J. C} \textbf{61} 471
(2009).
\bibitem{HDE5}  A. Sheykhi and M. Jamil \emph{Gen. Relativ. Gravit.} \textbf{43} 2661
(2011).
\bibitem{HDE6} M. Jamil and M. U. Farooq \emph{JCAP} \textbf{03} 001
(2010).
\bibitem{EHRDE1} L. N. Granda and A. Oliveros \emph{Phys. Lett. B}
\textbf{669} 275 (2008).
\bibitem{EHRDE2} Y. U. Fei and Z. Jing-Fei \emph{Commun. Theor. Phys.}
\textbf{59} 243 (2013).
\bibitem{Wicke} V. Acquaviva, C. Baccigalupi, S. M. Leach, A. R. Liddle and F.
Perrotta \emph{Phys. Rev. D} \textbf{71} 104025 (2005).



\end{thebibliography}
\end{document}